\documentclass[aps,prd,preprint,showpacs,showkeys,nofootinbib,superscriptaddress,floatfix,tightenlines]{revtex4-1}

\usepackage{amsmath}
\usepackage{amssymb,slashed}
\usepackage{dcolumn}	
\usepackage{bm}			
\usepackage{bbm} 		
\usepackage{multirow}
\usepackage{blkarray}
\usepackage[version=4]{mhchem}
\usepackage{feynmp}
\usepackage[usenames,dvipsnames]{xcolor}
\definecolor{hgreen}{rgb}{0,.3,0}
\definecolor{hred}{rgb}{.3,0,0}
\definecolor{hblue}{rgb}{0,0,.3}
\definecolor{LightGray}{gray}{0.95}
\usepackage[colorlinks=true,linkcolor=hblue,citecolor=hgreen,filecolor=hblue,
urlcolor=hred]{hyperref}
\makeatletter
\def\endfmffile{%
	\fmfcmd{\p@rcent\space the end.^^J%
		end.^^J%
		endinput;}%
	\if@fmfio
	\immediate\closeout\@outfmf
	\fi
	\ifnum\pdfshellescape>\z@
	\immediate\write18{mpost \thefmffile}%
	\fi}
\makeatother

\usepackage{hyperref,graphicx}           
\usepackage{pbox}

\usepackage{setspace}
\usepackage{caption}
\usepackage{subcaption}
\usepackage{xcolor}
\usepackage[utf8]{inputenc}

\newcommand{\beq}{\begin{equation}}
\newcommand{\eeq}{\end{equation}}
\newcommand\snowmass{\begin{center}\rule[-0.2in]{\hsize}{0.01in}\\\rule{\hsize}{0.01in}\\
\vskip 0.1in Submitted to the  Proceedings of the US Community Study\\ 
on the Future of Particle Physics (Snowmass 2021)\\ 
\rule{\hsize}{0.01in}\\\rule[+0.2in]{\hsize}{0.01in} \end{center}}

\begin{document}

\title{Exploring Dark Sector Portals with High Intensity Experiments}

\author{Brian Batell}
\affiliation{Pittsburgh Particle Physics, Astrophysics, and Cosmology Center, Department of Physics and Astronomy, University of Pittsburgh, Pittsburgh, Pennsylvania, USA\looseness=-1}
\author{Nikita Blinov}
\affiliation{Department of Physics and Astronomy, University of Victoria, Victoria, British Columbia
Canada\looseness=-1}
\author{Christopher Hearty}
\affiliation{University of British Columbia, Vancouver, British Columbia, 
Canada\looseness=-1}
\affiliation{Institute of Particle Physics (Canada), Victoria, British Columbia, 
Canada\looseness=-1}
\author{Robert McGehee}
\affiliation{Leinweber Center for Theoretical Physics, Department of Physics, University of Michigan, Ann Arbor, Michigan,
USA ~~~~~~~~~\looseness=-1}

\date{\today}
\begin{abstract}
A broad program of searches at high intensity experiments during the coming decade and beyond will sensitively probe new light mediator particles interacting through the minimal renormalizable vector, Higgs, and neutrino portals as well as higher-dimension axion-like particle portals. These portals may link the visible and dark sectors and play a critical role in many proposed solutions to some of the big open questions in particle physics and cosmology. 
In this whitepaper, we survey the theoretical and experimental progress, status, and prospects in the study of minimal dark sector portals. 
\end{abstract}



\maketitle
\snowmass
\tableofcontents

\section*{Executive Summary}
\label{sec.ExecSum}

The 2021 Snowmass Rare and Precision Frontier RF6 Topical Group, {\it Dark Sector Studies at High Intensities}, has solicited four {\it Big Ideas} white papers surveying the theoretical motivations and the experimental opportunities in the study of dark sectors. This white paper -- {\it Big Idea 2} -- covers the physics of {\bf minimal portal interactions}, including the renormalizable vector, Higgs, and neutrino portals, as well as minimal dimension-five, axion-like particle (ALP) portals with photon or gluon couplings. 
The focus is on minimal extensions of the Standard Model (SM) featuring a single new light mediator particle coupled through one of these portals. This implies that both the production of the mediator and its visible decay to SM particles occur due to the portal interaction. As we will examine in detail, there are a rich variety of exciting experimental opportunities to investigate the structure of the dark sector by producing and detecting such unstable mediator particles. We will review the current status and future prospects for exploring the minimal portals and highlight the connections with some of the big open questions in fundamental physics.
Three other RF6 {\it Big Ideas} white papers cover the complementary topics of dark matter production~\cite{Krnjaic:2022ozp}, rich and flavorful dark sectors~\cite{Harris:2022vnx}, and the experimental landscape for dark sector exploration at the intensity frontier~\cite{Ilten:2022lfq}.

The paradigm of a dark sector comprised of new SM singlet dark particles coupled to ordinary matter through a portal interaction is motivated on a variety of grounds. Dark sectors can resolve some of the outstanding mysteries in particle physics, including the dark matter puzzle, the dynamics underlying neutrino masses, baryogenesis, the hierarchy problem, the strong CP problem, and so forth.
From a bottom up perspective, portals provide a systematic effective field theory-based scheme for investigations of new light physics with very weak interactions. The dark sector framework has proven to be a versatile playground for exploring potential new physics explanations of an array of experimental anomalies. 
Dark sector research has bloomed over the past decade with the development of creative theoretical models, the conception of novel phenomenological strategies, and the proposal and implementation of innovative searches and novel experiments. 

The detailed properties of the mediator, including its mass, spin, and pattern of couplings to the visible sector, are of great interest from both theoretical and phenomenological perspectives. From the theory side, the gauge symmetries and field content of the SM impose tight constraints on the possible nature of the mediator and its couplings. On the other hand, these properties determine, to a significant extent, the possible phenomenological avenues that can be pursued to probe the mediator. In this light, the renormalizable vector, Higgs, and neutrino portals warrant special attention owing to their uniqueness and economy. These portals offer minimal ways to link gauge singlet scalar, fermion, or vector fields to the SM with sizable couplings at low energies. 
Beyond these three portals, the mediation between the visible and dark sectors can occur through higher-dimension portals. 
A well motivated and often studied case is a light ALP, with e.g., couplings to photons or gluons through dimension-five operators, whose mass is protected by a shift symmetry. While we will explore the renormalizable portals and minimal ALP portals here, it should, however, be emphasized that these are not the only ways of coupling a mediator to the SM. Other well motivated and phenomenologically distinctive possibilities will be examined in the other {\it Big Ideas} white papers.

Dark sectors are being pursued on multiple experimental fronts with a diverse set of search tactics. Electron and proton beam fixed target experiments 
with sensitive detectors covering ${\cal O}$(meter - kilometer) baselines 
provide excellent reach at low dark particle masses over a broad range of couplings.  Medium energy $e^+ e^-$ colliders/meson factories provide powerful sensitivity for moderate couplings both at low and intermediate masses. Precision studies of meson and lepton decays, including those at pion, kaon, $\eta^{(')}$, and muon facilities, offer interesting and in some cases unique coverage at low masses and small couplings. 
A diverse collection of existing and planned experiments at the LHC will be able to probe extensive regions of parameter space in a variety of dark sector models. 
Collectively, these experiments will utilize a wide array of search strategies, including bump-hunt searches for promptly decaying resonances, displaced vertex searches for dark particles with moderate lifetimes, searches for long lived particle decays to visible final states, and missing momentum searches in both collisions and rare decays.
These dedicated searches for dark sector particles are complemented by a variety of other probes in astrophysics and cosmology, precision measurements, and future energy frontier experiments. 

The minimal portals feature prominently in a variety of proposed solutions to the big questions in fundamental physics. 
These include a variety of motivated dark matter scenarios with novel cosmology and phenomenology. One generic example is secluded dark matter, in which heavier dark matter is thermally produced in the early universe via its annihilation to lighter mediator particles. Viable secluded DM models can be realized in any of the minimal portals. 
The requirement of thermalization in secluded scenarios imposes a lower bound on the portal coupling, offering an interesting target for high intensity and astrophysical probes. 
A variety of other interesting dark matter scenarios in which the mediator is the lightest dark sector state have been proposed, many of which can be correlated with specific regions of parameter space within the minimal portal models. The minimal portals may also play an important role in solutions to puzzles motivated by naturalness considerations. In particular, the Higgs portal is a critical ingredient in the relaxion solution to the hierarchy problem, while the vector portal is expected on general grounds and may have important consequences in the mirror Twin Higgs model, which tackles the little hierarchy problem. 
The ALP portal, and in particular the ALP-gluon interaction, is of course motivated by its connection to the Strong CP problem. The neutrino portal is likely to offer an explanation of the light SM neutrino masses and may also give rise to low-scale leptogenesis mechanisms. Furthermore, a light scalar interacting via the Higgs portal may also serve as the inflaton. This white paper will spotlight the myriad connections between the minimal portal models and the potential answers to the big questions in particle physics and cosmology. 

The theoretical ideas and experimental approaches featured in this whitepaper have important synergies and complementarity with other efforts across the rare and precision, energy, cosmic, and neutrino frontiers. At the energy frontier, the LHC and future high energy colliders will be able to probe heavier mediators with larger couplings. Distinct experimental opportunities are also available at the energy frontier, including e.g., exotic Higgs decays and precision measurements of Higgs couplings and electroweak observables. At the cosmic frontier, a suite of new direct detection experiments will directly search for halo DM through its scattering, while an array of astrophysical observations can indirectly search for signatures of DM annihilation. In particular, these direct and indirect searches provide sensitivity to DM that is heavier than the mediator and are therefore highly complementary to the direct searches for the visibly decaying mediators at intensity frontier experiments highlighted in this whitepaper. There is also an exciting interplay with activities in the neutrino frontier. Dark sector mediators can be sought at accelerator- and reactor-based neutrino experiments and have also been invoked in a variety of potential BSM explanations for various experimental anomalies in the neutrino sector (e.g., the MiniBooNE low energy excess of electron like events). Within the RF6 topical group, two additional big ideas whitepapers will cover various complementary aspects of the dark sectors physics program. Big Idea 1 focuses on dark matter production and covers invisible or semi-visible mediator decays to dark matter and search strategies leveraging missing energy/momentum or DM re-scattering. Big Idea 3 explores rich and flavorful dark sectors, highlighting mediators with novel flavor structure and more complex dark sectors, which may be anticipated in a more complete models.

\begin{figure}[h]
\begin{center}
\includegraphics[width=0.85\textwidth]{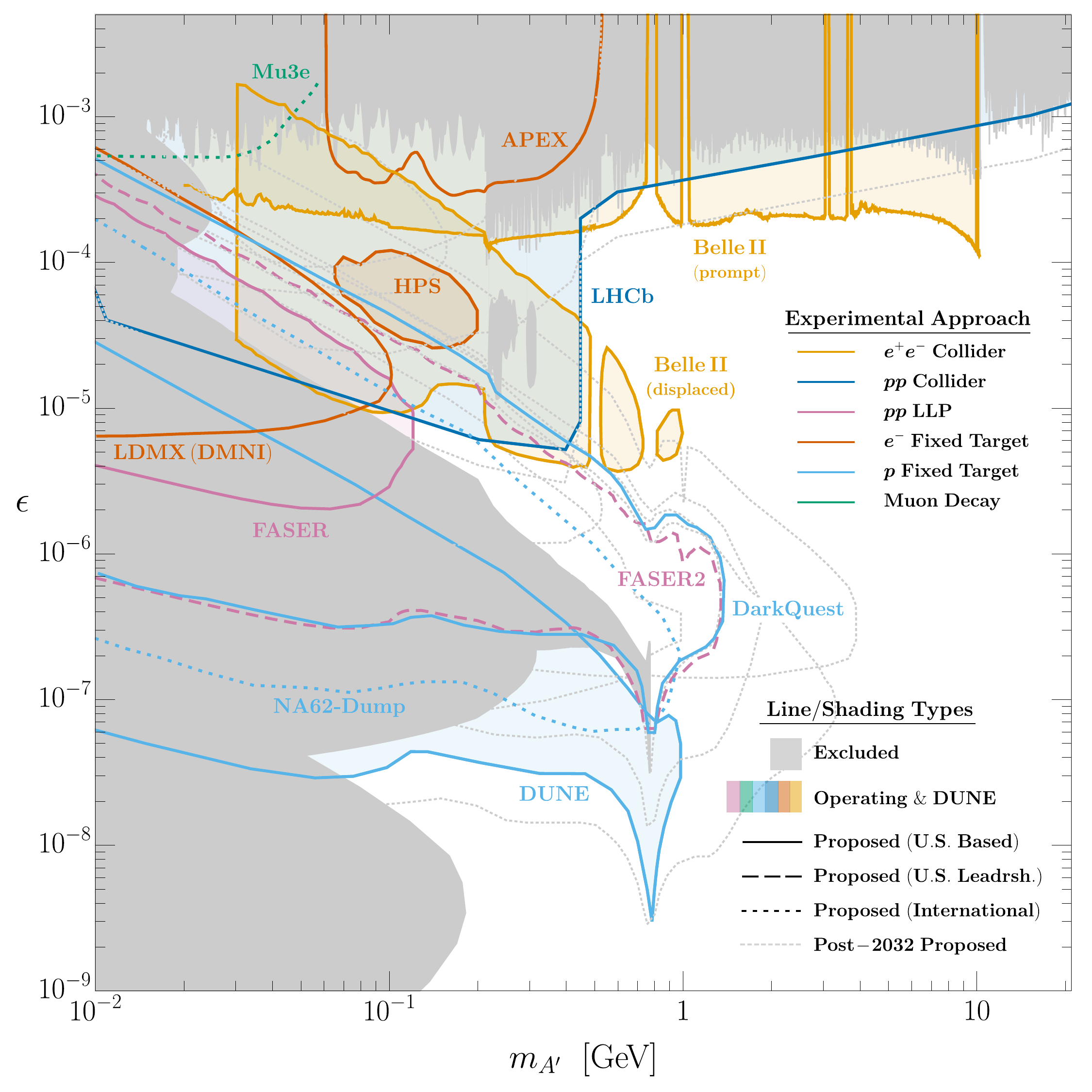}
\end{center}
\caption{
Near-term and future opportunities to search for visibly decaying massive dark photons interacting through the kinetic mixing vector portal displayed in the dark photon mass $(m_{A'})$ -- kinetic mixing  $(\epsilon)$ parameter space. Constraints from past experiments (gray shaded regions) and projected sensitivities from operating experiments and DUNE (colored shaded regions), proposed near-term (pre-2032) experiments based in the U.S. including Dark Matter New Initiatives (DMNI) supported experiments (solid colored lines), proposed near-term (pre-2032) experiments based internationally and having significant U.S. leadership (dashed colored lines), proposed near-term (pre-2032) international projects (dotted colored lines), and proposed future (post 2032) experiments (dotted gray lines) are shown; see also Figure~\ref{fig:vector} for another version of this plot with all future experiments labeled. Line coloring indicates the key experimental approach used ($e^+ e^-$ collider, $pp$ collider, LHC LLP detector, electron fixed target, proton fixed target, muon decay), highlighting one aspect of the complementarity between different facilities/experiments. Collectively, these experiments are poised to cover large regions of open dark photon and thermal dark matter parameter space. 
}
\label{fig:Vector-Summary}
\end{figure}

\begin{figure}[h]
\begin{center}
\includegraphics[width=0.85\textwidth]{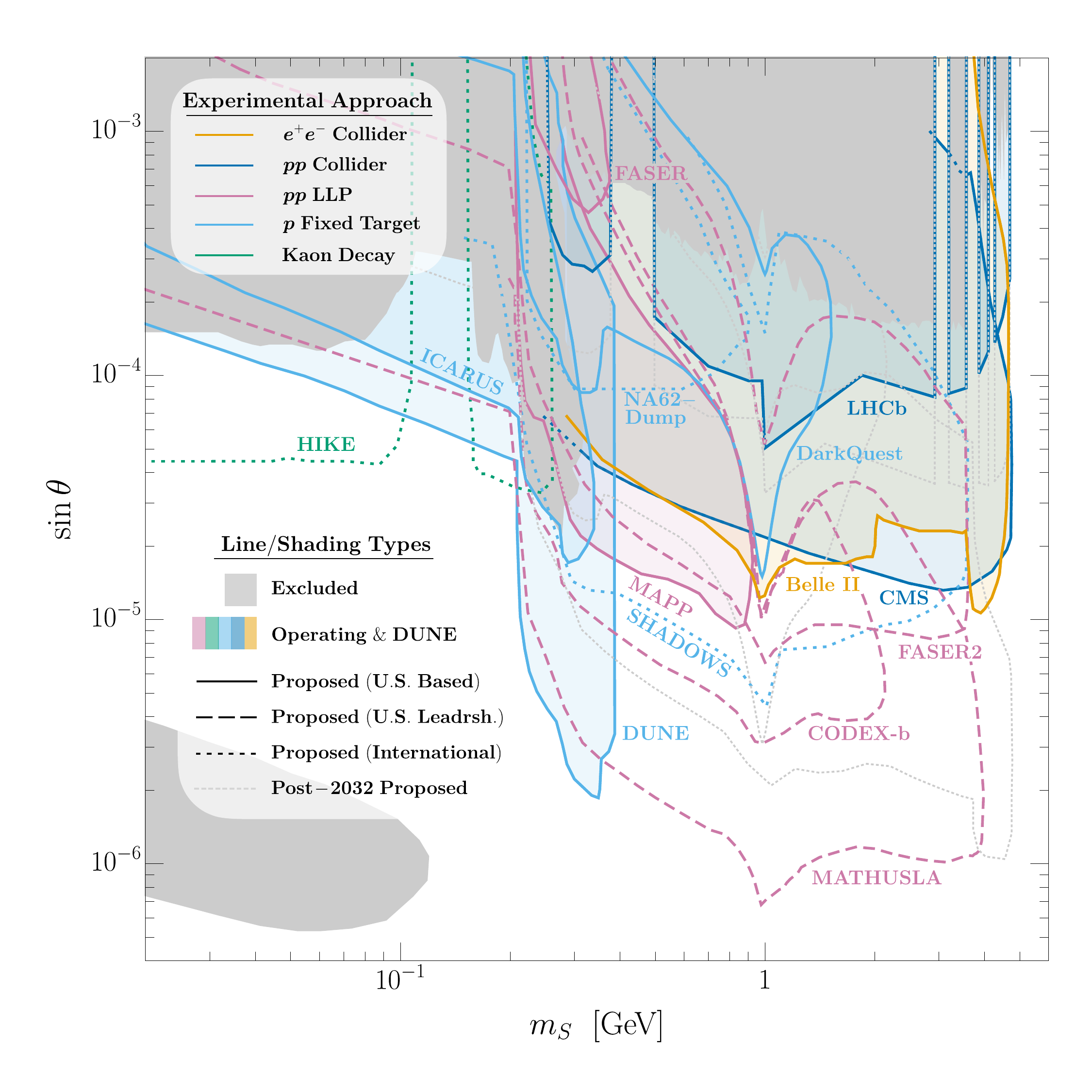}
\end{center}
\caption{
Near-term and future opportunities to search for visibly decaying massive dark scalars interacting through the Higgs portal displayed in the scalar mass $(m_{S})$ -- mixing angle $(\sin\theta)$ parameter space. Constraints from past experiments (gray shaded regions) and projected sensitivities from operating experiments and DUNE (colored shaded regions), proposed near-term (pre-2032) experiments based (solid colored lines), proposed near-term (pre-2032) experiments based internationally and having significant U.S. leadership (dashed colored lines), proposed near-term (pre-2032) international projects (dotted colored lines), and proposed future (post 2032) experiments (dotted gray lines) are shown; see also Figure~\ref{fig:scalar} for another version of this plot with all future experiments labeled. Line coloring indicates the key experimental approach used ($e^+ e^-$ collider, $pp$ collider, LHC LLP detector, proton fixed target, kaon decay), highlighting one aspect of the complementarity between different facilities/experiments. Collectively, these experiments are poised to cover large regions of open dark scalar and thermal dark matter parameter space.
}
\label{fig:Scalar-Summary}
\end{figure}

As illustrative examples, Figs.~\ref{fig:Vector-Summary} and \ref{fig:Scalar-Summary} present the near-term and future opportunities to probe the minimal vector portal and Higgs portal models, respectively.
Dark photons and Higgs portal scalars are well-motivated dark sector benchmarks, may serve as a mediators to dark matter, and appear in a variety of UV models addressing big open questions in particle physics. 
A combination of operating, fully or partially funded, and proposed near term and future experiments will be able to search for dark photons over a broad range of currently unconstrained parameter space. As this example illustrates, and as will be highlighted in this whitepaper with specific case studies, there is great opportunity to explore vast uncharted parameter space and investigate the structure of the dark sector during the next decade and beyond.

The dark sector science program has developed significantly during the last decade. 
Milestones in this trajectory can be seen in past community studies, including the Dark Sectors 2016 Workshop~\cite{Alexander:2016aln} and the US Cosmic Visions New Ideas in Dark Matter 2017~\cite{Battaglieri:2017aum} and the CERN Physics Beyond Colliders~\cite{Beacham:2019nyx}. Furthermore, the U.S. Department of Energy (DOE) Office of Science Dark Matter New Initiatives (DMNI) Basic Research Needs (BRN) report~\cite{BRNreport} has recently highlighted the importance of MeV-GeV scale dark sector studies as a Priority Research Direction: ``Create and detect dark matter particles below the proton mass and associated forces, leveraging DOE accelerators that produce beams of energetic particles.'' In particular, the need to search for and study visible signatures of dark sectors is emphasized through Thrust 2: ``Explore the structure of the dark sector by producing and detecting unstable dark particles.'' Already, the DMNI initiative is supporting CCM200 at Los Alamos’ LANSCE proton beam and LDMX at SLAC’s LESA electron beamline. It will be critical to continue supporting and expand these efforts going forward. 

Realizing the broad objectives of the dark sector science program, as spotlighted by the three Big Ideas, necessitates efforts and investments in several primary directions, including harnessing the potential of existing large-scale multi-purpose detectors, supporting dedicated small-scale experiments and facilities with high-intensity beams, and promoting advances in dark sector theory and fostering the community of dark sector theorists. Support for these projects and research activities will facilitate a broad and high-impact dark sector physics program during the next decade, with US scientists and institutions providing key leadership in this effort.

\section{Introduction}
\label{sec:Intro}

The SM of particle physics is an extraordinarily successful theory, correctly describing familiar matter and forces down to length scales of at least $10^{-18}$ m and playing an essential role in our understanding of the history of the Universe. 
The spectacular successes of the SM are matched only by its striking failures to provide answers to a handful of open questions raised by both empirical observations and conceptual mysteries. What is dark matter? What dynamics is responsible for neutrino masses? How is the matter-antimatter asymmetry generated? What physics underlies the Higgs sector and sets the weak scale? Why is CP conserved by the strong interactions? These and other questions strongly motivate explorations of physics beyond the SM.

On general grounds, the new dynamics addressing these questions could manifest in several ways. First, if the new degrees of freedom are significantly heavier than the weak scale, one can resort to searches for anomalous phenomena or rare processes with precision measurements. Another possibility involves new states charged under the SM gauge symmetries with masses near the weak scale, which can be directly probed by experiments at the energy frontier. Finally, there may be a {\it dark sector} (or hidden sector, etc.) containing new SM gauge singlet states. The dark sector states may have masses well below the weak scale and communicate weakly with the visible sector through a {\it portal} interaction linking gauge invariant SM operators to a mediator. 
Speculations regarding dark sectors have gained significant traction over the last decade due to their possible connections with the big questions mentioned above as well as their rich phenomenology.

The dark sector could be quite minimal, consisting of one or a few new states, or it could be as rich as the SM, containing new dark forces, dark matter states, and dark Higgs fields while exhibiting novel phenomena such as confinement or spontaneous symmetry breaking. While it is important to examine the theoretical and phenomenological implications of rich dark sectors, including their potential role in addressing the big questions such as e.g., the dark matter puzzle, there is also merit in exploring the physics of minimal portals, i.e., a single new mediator and portal coupling in isolation. From a bottom-up effective field theory perspective, the portal concept provides a natural point of departure for the systematic exploration of new light weakly coupled physics. As is well-known, as a consequence of the SM gauge symmetries and field content, there are just three options for renormalizable portals: the vector portal, the scalar portal, and the neutrino portal. At the dimension five level and higher, a variety of portal couplings are allowed, including in particular the ALP-portal couplings to photons and to gluons. ALPs may be naturally light as a consequence of a shift symmetry, and these particular couplings are motivated by their appearance in solutions to the Strong CP problem. Another clear motivation for considering minimal portals is that the phenomenology of the mediator may naturally map on to more complex dark sector models in the case that the mediator is the lightest new state in the theory. 

With this motivation, the scope of this 2021 Snowmass RF6 {\it Big Idea 2} white paper is on the physics of the mediator in the minimal vector, Higgs, neutrino, and ALP portal models. These simple, well-motivated benchmark models predict a rich variety of phenomena and motivate an expansive set of new searches and dedicated experiments at the intensity frontier, complementing other probes at the cosmic and energy frontiers. In these models, the single portal coupling dictates both the production channels of the mediator in collisions or rare decays of SM particles as well as its visible decay modes to SM final states. Two other RF6 {\it Big Ideas} white papers cover the complementary topics of dark matter production and rich and flavorful dark sectors. 

Next, in Section~\ref{sec:Exp}, we survey the various experimental approaches available and the specific existing or proposed experiments that can probe the minimal portals. These include electron and proton beam fixed target experiments, medium energy $e^+ e^-$ colliders/meson factories, pion, kaon, 
$\eta^{(')}$, muon sources, and a variety of experiments at the LHC. 

Following this, in Section~\ref{sec:models} we introduce the minimal portal models, describe their basic properties and interactions, discuss the future prospects for probing these models at a diverse collection of experiments in the coming years, and also highlight how these minimal portals may feature in more complete dark sector models that address the big questions in fundamental physics. 

\section{Experimental approaches}
\label{sec:Exp}

The searches covered by this white paper involve a mediator decaying to a pair of SM particles. The general search strategies depend on the mass of the mediator and its couplings to the SM,  which dictate its possible production modes, decay channels, and lifetime.

Prompt decays, where the decay vertex is experimentally indistinguishable from the production point, have large SM backgrounds,  e.g., $\gamma^* \to \ell^+ \ell^-$ for dilepton resonance signals. On an event-by-event basis, these are indistinguishable from signal. Instead, a signal is identified as a narrow peak on a smooth mass distribution from the SM background. Sensitivity improves with better mass resolution and large data sets. 
For long lifetimes, several to hundreds of meters in the laboratory, large amounts of shielding can substantially reduce SM backgrounds. Given the dependence of the lifetime on mass and coupling, these searches are necessarily focused on low masses and small couplings. The boost of the mediator in the laboratory frame, which plays a large role in determining the range of relevant couplings, depends strongly on the energy of the beam used to produce the mediator, as well as on the mediator mass. 

Between these two cases, precision vertex detectors are used to distinguish prompt SM backgrounds from displaced signals. A detailed understanding of the location of detector materials is required to distinguish backgrounds from SM particles interacting in the material. Searches in electron pair final states are particularly affected by photon conversions. SM particles with comparable lifetimes, such as $K^0_s$ mesons and $\Lambda$ baryons, are also a consideration. 

The different experiments that have produced results or projections can be sorted into the following categories. Details on each experiment and facility along with references are in the RF6 facilities white paper~\cite{Ilten:2022lfq}.

\subsection{\texorpdfstring{$e^+e^-$}{e+ e-} colliders}

Experiments at $e^+e^-$ colliders can perform a wide range of searches for different mediators, with either prompt or intermediate lifetimes. The production mechanism can be either $e^+e^-$ annihilation, via coupling to photons, or, for facilities operating at the $\Upsilon(4S)$ resonance, $B$ meson decay. Searches are enhanced by the large acceptance and nearly-hermetic design of the detectors. Current facilities have modest center of mass energies $\mathcal{O}(10)$\,GeV, but future colliders could significantly extend the mass reach. 

Relevant experiments that have completed data collection include BaBar and KLOE. BaBar collected 500\,fb$^{-1}$ of data at the PEP-II asymmetric $e^+e^-$ collider at SLAC from 1999--2008. KLOE operated at the DA$\Phi$NE collider from 2001--2006; the upgraded KLOE-2 collected data from 2014--2018. DA$\Phi$NE operates at the $\phi$ resonance. 

Belle II is currently collecting data at the SuperKEKB collider at KEK, in Tsukuba, Japan. It is scheduled to record 50\,ab$^{-1}$ at the $\Upsilon(4S)$ over the next decade. Proposed future colliders include the International Linear Collider (ILC), the Future Circular Collider (FCC-ee), the Circular Electron-Positron Collider (CEPC), and the Cool Copper Collider (C3). 

\subsection{Proton or electron beam dumps}

Beam dump experiments look for the appearance of visible decay products in a detector separated from the production target by sufficient shielding to reduce SM backgrounds to 
manageable levels. 
They have very high intensities and probe longer lifetimes, which gives them sensitivity to small couplings. The typical mass reach is less than a few GeV$/c^2$. 

The results of older electron beam dump experiments E141, E137, E774, KEK, and Orsay have been recast into limits on dark photon production. The dark photon is produced via bremsstrahlung, $eZ \to eZ A^\prime$.

Results from proton beam dump experiments have also been recast as dark photon or scalar limits. $\nu-$CAL I uses proton bremsstrahlung and $\pi^0$ decay; CHARM, NOMAD, and PS191 all rely on $\pi^0$ decay.

DarkQuest, an upgrade of the existing SeaQuest/SpinQuest experiment, will use 120\,GeV protons from the Fermilab main injector to search for dark photons and scalars. Production mechanisms include Drell-Yan, meson decay, and proton bremsstrahlung. Two proposals would use 400\,GeV protons extracted from the CERN SPS. SHiP will search for a wide range of feebly interacting particles. SHADOWS, a somewhat smaller facility, will be located off-axis adjacent to NA62. Both will use sets of dipole magnets to sweep away muons and other charged particle backgrounds. 

\subsection{Electron beam fixed target experiments}

Electron beam fixed target experiments produce dark photons or scalars in bremsstrahlung and search for the subsequent decay to $e^+e^-$. For prompt decays, the search is for a narrow peak on a large background from $\gamma^* \to e^+e^-$.  

The HPS experiment at JLAB looks for displaced decays using a high-resolution silicon vertex tracker. Data has been collected at a beam energy of 4.55\,GeV; additional data sets will be collected with beam energies as high as 6\,GeV.

NA64 is a fixed target experiment using a 100--150\,GeV electron beam from the CERN SPS to search for a variety of dark sector particles in both visible and invisible modes. It features an active target separated from a deep calorimeter by a decay volume and veto and tracking elements. 

It should be noted that one can also probe dark sector particles with positron beam fixed target experiments; see, e.g., Refs.~\cite{Rachek:2017gdc,PADME:2021vjp,Duarte:2022feb}.

\subsection{Large Hadron Collider}

Dark sector particles are produced in proton-proton collisions via meson decay, dark photon mixing with $\rho$, $\omega$, and $\phi$ mesons, or Drell-Yan ($q \bar q$ annihilation). LHCb and CMS have searched for dark photons decaying to muon pairs. LHCb uses both prompt and displaced vertices, and intends to add the $e^+e^-$ final state. 

FASER is located 480\,m downstream of the ATLAS collision point. It exploits the large cross section in the forward direction and the large boost of light particles to search for long-lived dark particles. It will collect first data starting summer 2022. 
MoEDEL-MAPP, located 100\,m from the LHCb interaction point, will search for minicharged and other feebly-interacting particles starting in 2023.
Other proposed long-lived particle experiments at the LHC include CODEX-b and MATHUSLA. Both are at wide angles, giving it access to production in the decay of heavier particles. The proposed Forward Physics Facility~\cite{Anchordoqui:2021ghd,Feng:2022inv} would house a variety of experiments sensitive to new physics, including FASER2, an upgrade of FASER. 

\subsection{Meson and lepton facilities}

Dark sector mediators can be produced in rare decays of Kaon or eta mesons, or of muons. Searches exploit the large data sets available, and the narrow width of the mother particle. NA48/2 has searched for prompt decay of dark photons to $e^+e^-$ in $\pi^0$ decay. The NA62 experiment plans a wide range of searches, including prompt and displaced electron pairs, muon pairs, and photon pairs, and searches for long-lived mediators via missing mass techniques.  

REDTOP is a proposed very-high statistics $\eta$ and $\eta^\prime$ facility that would be sensitive to a variety of dark sector particles through searches in lepton pair final states. 

Mu3e, located at the Paul Scherrer Institute (PSI), is principally a search for the (forbidden) decay $\mu^+\to e^+e^-e^+$, but will also have sensitivity to dark photon decays to $e^+e^-$. Also located at PSI is PIONEER. It will primarily study lepton flavor universality in charged pion decay, but will also have sensitivity to heavy neutral leptons.

\section{Minimal Portal Case Studies}
\label{sec:models}

We now turn our attention to minimal portal models, including the vector portal, Higgs portal, neutrino portal, and ALP portal with photon or gluon couplings. We discuss the interactions of the mediator with the SM and its essential properties concerning its production, decay, and lifetime that enter into phenomenological considerations. We then consider the current status and constraints on the mass-coupling parameter space in each model, and also highlight the future prospects at a variety of existing or proposed experiments. Finally, we discuss how these minimal portals, within the context of more complete dark sector models, may play a role in addressing the outstanding questions in particle physics.

\subsection{Vector Portal}
\label{subsec:vector-portal}

\begin{figure}[h]
\begin{center}
\includegraphics[width=0.9\textwidth]{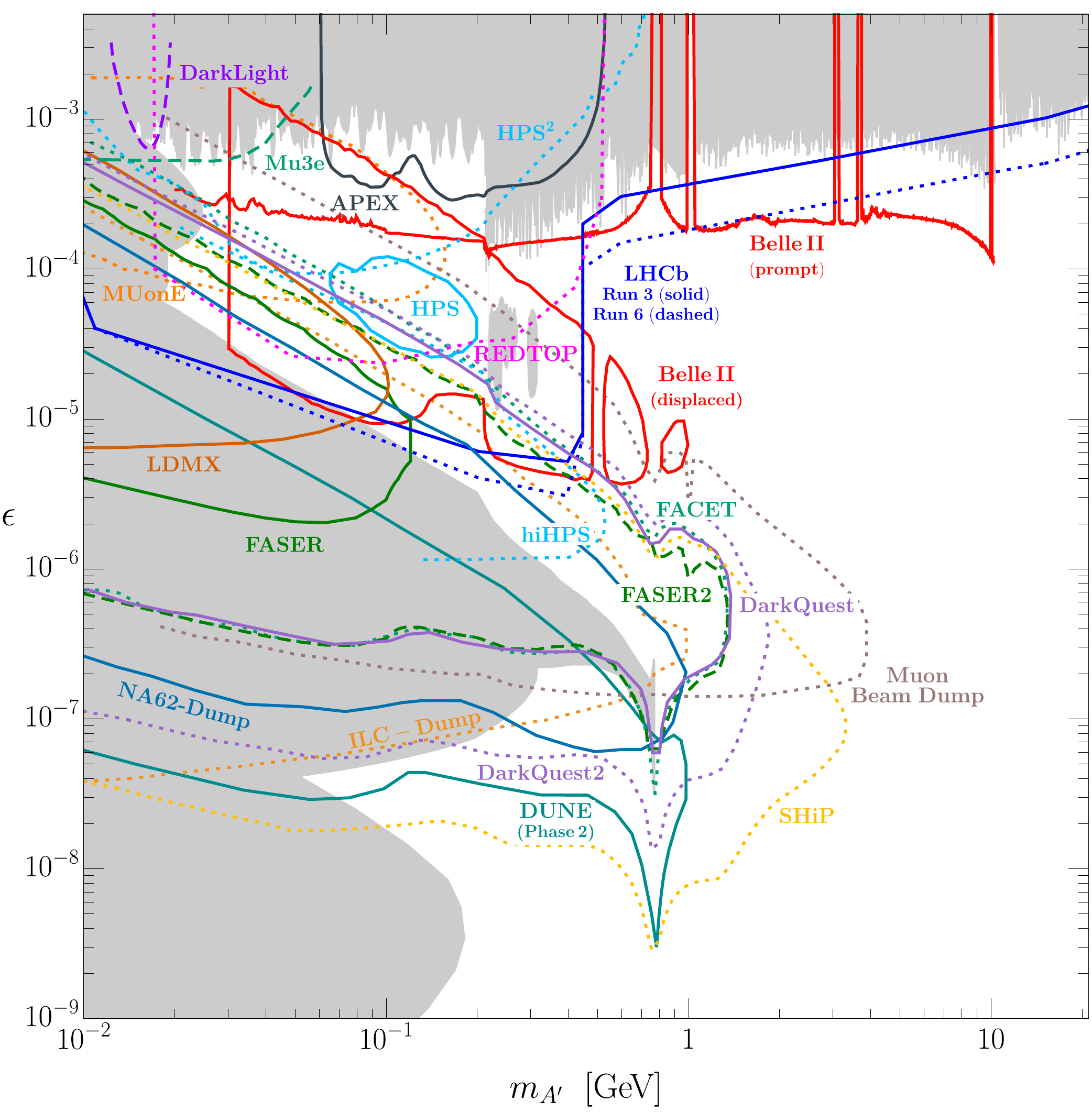}
\end{center}
\caption{\textit{Minimal vector portal model}. Existing constraints (gray shaded region) and future experimental projections (colored contours) are shown in the $m_{A'}-\epsilon$ plane. The existing bounds are from 
Refs.~\cite{Blumlein:2011mv,Blumlein:2013cua,Tsai:2019buq,Bjorken:2009mm,Andreas:2012mt,Bjorken:1988as,Bross:1989mp,Riordan:1987aw,Davier:1989wz,Banerjee:2019hmi,Batley:2015lha,Merkel:2014avp,Abrahamyan:2011gv,KLOE-2:2018kqf,LHCb-PAPER-2017-038,Aaij:2019bvg,Ablikim:2017aab,Lees:2014xha,Sirunyan:2019wqq,Curtin:2014cca}.
Also shown are projections from a number of existing and proposed future experiments, including 
DUNE~\cite{Berryman:2019dme},
Belle II~\cite{Belle-II:2018jsg,Ferber:2022ewf,Bandyopadhyay:2022klg},
LHCb~\cite{Craik:2022riw},
FASER and FASER2~\cite{Feng:2017uoz,FASER:2018eoc}, 
HPS~\cite{Baltzell:2022rpd},
NA62-Dump~\cite{Beacham:2019nyx},
LDMX~\cite{Berlin:2018bsc},
DarkQuest~\cite{Berlin:2018pwi},
APEX~\cite{Essig:2010xa},
Mu3e~\cite{Echenard:2014lma},
DarkLight~\cite{DarkLight},
FACET~\cite{Cerci:2021nlb},
REDTOP~\cite{REDTOP:2022slw},
MUonE~\cite{Galon:2022xcl},
SHiP~\cite{SHiP:2020vbd}, an
ILC beam dump experiment~\cite{Kanemura:2015cxa,Asai:2021ehn}, and
a muon beam dump experiment~\cite{Cesarotti:2022ttv}.
}
\label{fig:vector}
\end{figure}

We will first consider the vector portal, which couples a new dark $U(1)$ gauge boson $A'_\mu$, commonly referred to as a dark photon, to the hypercharge gauge boson via the kinetic mixing operator,
\begin{equation}
{\cal L} \supset \frac{\epsilon }{2 \cos \theta_{W}} F'_{\mu\nu} \, B^{\mu\nu}.
\label{eq:vector-portal}
\end{equation}
Here $F'_{\mu\nu}$ ($B_{\mu\nu}$) is the dark photon (hypercharge) field strength tensor,  $\epsilon$ is the kinetic mixing parameter, and $\theta_W$ is the weak mixing angle. 
The kinetic mixing operator, Eq.~(\ref{eq:vector-portal}), is allowed at the renormalizable level by all symmetries of the theory, may be generated radiatively~\cite{Holdom:1985ag}, or have important connections to UV physics, and a broad range of values for $\epsilon$ are well motivated theoretically~\cite{Holdom:1985ag,Dienes:1996zr,Arkani-Hamed:2008kxc,Koren:2019iuv,Gherghetta:2019coi}. 
We assume here that the dark photon obtains a mass $m_{A'}$ via a dark Higgs or Stueckelberg mechanism, such that the minimal vector portal parameter space is $(m_{A'}, \epsilon)$.

Dark photons that are significantly lighter than the $Z$ boson dominantly couple in the physical basis to electrically charged particles with interaction strength suppressed by $\epsilon$. 
As such, dark photons will decay democratically to all kinematically accessible electrically charged particles. The dark photon decay length scales parametrically as $c \tau_{A'} \sim (\epsilon^{2} m_{A'})^{-1}$, and may be prompt for moderate kinetic mixing strength ($\epsilon \gtrsim {\cal O}(10^{-3})$) or displaced/ macroscopic for smaller values of $\epsilon$. As it couples with similar strength to both charged leptons and quarks, there are a variety of promising experimental venues for dark photon searches in high intensity facilities utilizing hadrons, electrons, and muons. In particular, dark photons can be copiously produced and probed through their visible decays at high-luminosity $e^+ e^-$ colliders, electron and proton beam fixed target experiments (employing bump hunt, displaced vertex, and LLP decay searches), meson factories, and the LHC. 
Figure~\ref{fig:vector} displays the current bounds and future sensitivity projections from a variety of experiments in the $m_{A'}- \epsilon$ parameter space.

Visibly-decaying dark photons detectable at near-future experiments are motivated by many solutions to the big questions including the nature of DM. For secluded DM models in which DM annihilates to lighter dark-sector states \cite{Pospelov:2007mp,Arkani-Hamed:2008hhe}, the requirement of thermal equilibrium for freeze-out predicts one such target \cite{Evans:2017kti}. In Forbidden Dark Matter scenarios, DM instead annihilates to heavier states, such as visibly-decaying dark photons with masses and kinetic mixings in range of future searches \cite{DAgnolo:2015ujb}. Going to heavier dark photons motivates \emph{Not} Forbidden Dark Matter, where the mediator is too heavy to permit even suppressed $2 \to 2$ annihilations, so $3 \to 2$ annihilations determine the DM relic abundance. Again, using the vector portal is a natural simple choice and some of the viable parameter space which reproduces the relic abundance will be probed at future experiments \cite{Cline:2017tka}. Resonant Dark Matter models consider the possibility of resonantly enhancing DM annihilations through the vector portal to evade cosmological constraints. Interestingly, smaller tuning of the resonance results in a more predictive target for dark photon searches \cite{Bernreuther:2020koj}. The resonance may not be in the vector portal itself, but rather in DM self interactions to alleviate small-scale structure issues. The resulting Resonant Self-Interacting Dark Matter may still use the vector portal for entropy transfer and again, less fine tuning motivates a narrower region for future dark photon searches to probe \cite{Tsai:2020vpi}. Some Strongly Interacting Massive Particle scenarios, in which $3 \to 2$ self annihilations set the DM relic abundance, also use the vector portal to transfer excess entropy out of the dark sector. The resulting viable parameter space which explains the DM relic abundance while evading all other constraints can be quite predictive in the $m_{A'}-\epsilon$ plane and may be imminently probed \cite{Choi:2017zww}. 

The vector portal has also been present in answers to the naturalness and baryon asymmetry questions. For example, the Mirror Twin Higgs paradigm always contains the vector portal since it proposes a copy of the SM matter and gauge content to solve the little hierarchy problem~\cite{Chacko:2005pe}. One such model alleviates otherwise-expected cosmological tensions and provides an imminently-testable, well-motivated benchmark for many future searches due to naturalness considerations \cite{Harigaya:2019shz}. Another realizes a Strongly Interacting Massive Particle DM scenario within the Mirror Twin Higgs framework and predicts visibly-decaying ``twin'' photons in the process \cite{Hochberg:2018vdo}. As for the baryon asymmetry, one model provides an explanation and a simultaneous asymmetric dark matter candidate using a dark sector similar to the Mirror Twin Higgs. It thus has a vector portal which must efficiently transfer entropy prior to BBN and, due to the precisely predicted ratio between SM and dark baryon asymmetries, requires $10 \text{ MeV} \lesssim m_{A'} \lesssim 300 \text{ MeV}$, which will be probed by many visibly-decaying dark photon searches \cite{Hall:2019rld}.

\subsection{Higgs Portal}
\label{subsec:higgs-portal}

\begin{figure}[tbh]
\begin{center}
\includegraphics[width=0.9\textwidth]{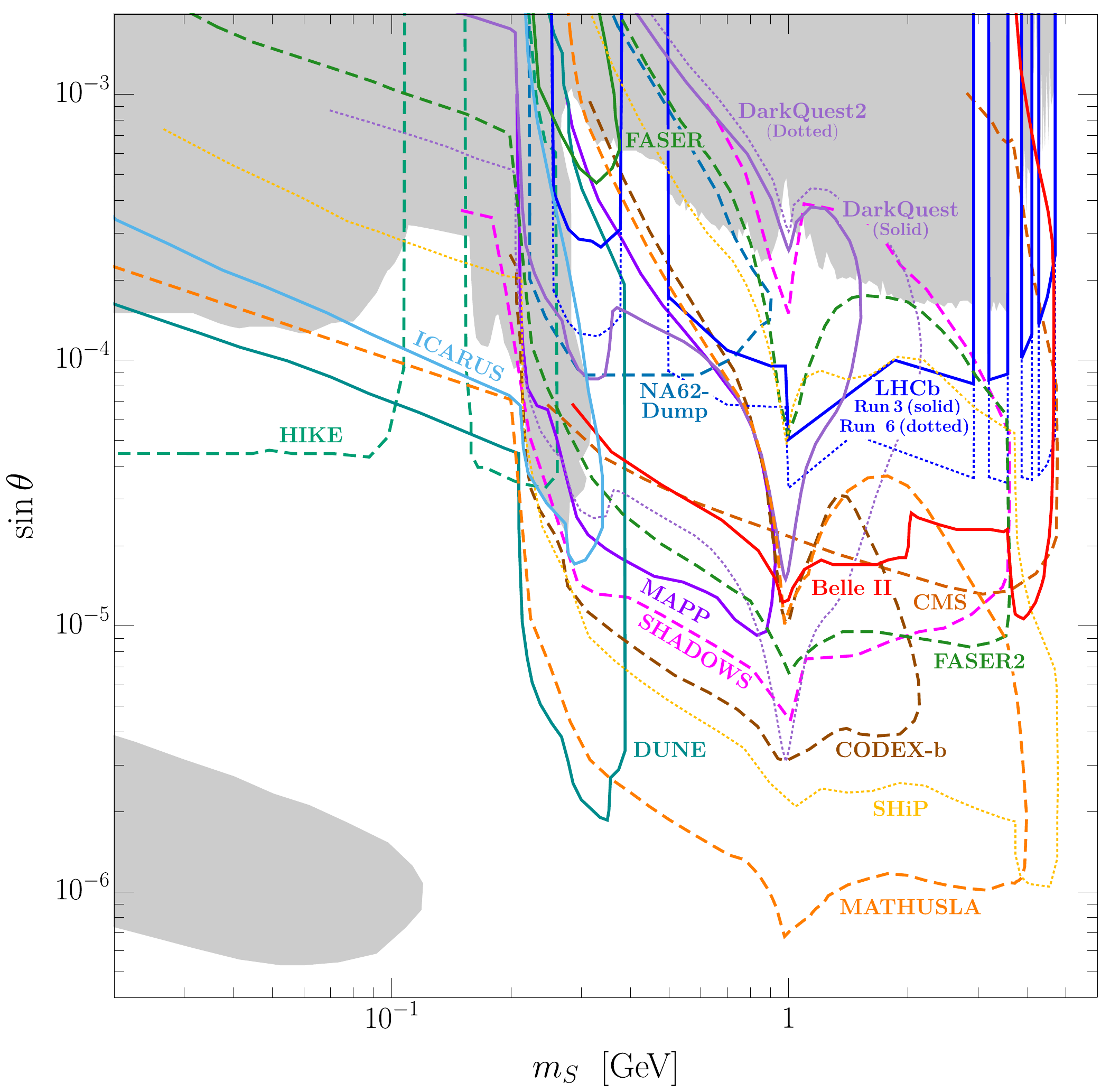}
\end{center}
\caption{\textit{Minimal Higgs portal model}. 
Existing constraints (gray shaded region) and future experimental projections (colored contours) are shown in the $m_{S}-\sin\theta$ plane. The existing bounds are from 
Refs.
\cite{Frugiuele:2018coc,Aaij:2015tna,Aaij:2016qsm,LSND:1997vqj,LSND:2001aii,Foroughi-Abari:2020gju,Bernardi:1985ny,Bernardi:1987ek,Gorbunov:2021ccu,MicroBooNE:2021usw,Bergsma:1985qz,Winkler:2018qyg,Egana-Ugrinovic:2019wzj,Goudzovski:2022vbt,Artamonov:2008qb,Artamonov:2009sz,NA62:2021zjw,NA62:2020xlg,NA62:2020pwi}.
Also shown are projections from a number of existing and proposed future experiments, including 
ICARUS~\cite{Batell:2019nwo},
DUNE~\cite{Berryman:2019dme},
Belle II~\cite{Kachanovich:2020yhi,Filimonova:2019tuy}, 
LHCb~\cite{Craik:2022riw},
CMS~\cite{Evans:2020aqs},
DarkQuest~\cite{Batell:2020vqn,Berlin:2018pwi},
FASER and FASER2~\cite{Feng:2017vli},
MoEDAL-MAPP~\cite{Pinfold:2791293,MoEDAL},
CODEXb~\cite{Aielli:2022awh},
MATHUSLA~\cite{Curtin:2018mvb},
NA62-Dump~\cite{Beacham:2019nyx},
SHADOWS~\cite{SHADOWS}, 
HIKE~\cite{SHADOWS},
and SHiP~\cite{Alekhin:2015byh}. }
\label{fig:scalar}
\end{figure}

Next, we turn to the Higgs portal, which couples the gauge invariant Higgs mass operator $H^\dag H$ to a new gauge singlet scalar particle $S$. There are two allowed renormalizable couplings in general, 
\begin{equation}
-{\cal L} \supset (A \, S +\lambda \, S^2) \, H^\dag H  .
\label{eq:Higgs-portal}
\end{equation}
where $A$ has dimensions of energy and $\lambda$ is dimensionless. 
If $A$ is non-vanishing, or even if $A = 0$ and the scalar potential leads to a vacuum expectation value for $S$, the dark scalar will acquire a small mass mixing with the Higgs boson.
The phenomenology of the Higgs portal is thus broadly described by two parameters: the physical mass $m_S$ of the dark scalar and the scalar-Higgs mixing angle $\theta$. The dark scalar inherits the interactions of the Higgs to SM particles, with a coupling suppressed by $\sin\theta$. As such it may be abundantly produced through flavor-changing meson decays, e.g. $K \rightarrow \pi S$, $B\rightarrow K S$, etc, and furthermore will dominantly decay to the heaviest kinematically accessible final states. For $m_S < 2 m_\mu$, the scalar decays to dielectrons with a naturally long lifetime, and is probed in beam dump experiments, dedicated LLP LHC detectors, and in rare kaon decays with missing energy. At higher masses above the dimuon threshold, the dark scalar may additionaly be sought in experiments studying rare $B$ meson decays. 
Figure~\ref{fig:scalar} displays the current experimental bounds and expected reach from a variety of
experiments in the dark scalar parameter space.
It is important to note that there may be additional signatures in models with a sizable $hSS$ trilinear coupling (e.g., exotic Higgs boson decays); see e.g., Refs.~\cite{Beacham:2019nyx,Frugiuele:2018coc} for further studies of this possibility. 

The Higgs portal is also motivated by simple solutions to many of the big questions, including the nature of DM. It appears in predictive models of secluded DM \cite{Pospelov:2007mp} as well as light thermal DM \cite{Krnjaic:2015mbs}. Some such secluded DM models can produce exciting targets for future experiments due to thermal equilibrium requirements \cite{Evans:2017kti}, as can models of SIMP DM in which the Higgs portal is an important necessity again \cite{Choi:2017zww}. The Higgs portal also naturally appears in solutions to the electroweak hierarchy problem. The relaxion, a scalar which dynamically reduces the mass of the Higgs during inflation, may naturally mix with the Higgs at near-detectable levels \cite{Flacke:2016szy,Frugiuele:2018coc}. Another light, cosmologically-evolving scalar is the inflaton itself, which may also mix with the Higgs and lead to interesting signals \cite{Bezrukov:2009yw}.

\subsection{Neutrino Portal}
\label{subsec:neutrino-portal}

The neutrino portal refers to the coupling of the gauge invariant operator $LH$ formed of the lepton and Higgs doublets to a gauge singlet fermion $N$, 
\begin{equation}\label{eq:NeutrinoMassLagrangian}
    \mathcal{L} \supset -y^{\alpha} L_\alpha H N +{\rm h.c.},
\end{equation}
where $y^\alpha$ is a Yukawa coupling with $\alpha = e,\, \mu,\, \tau$.
We refer to $N$ as a heavy neutral lepton (HNL). Following electroweak symmetry breaking, the HNLs mix with the SM neutrinos, inheriting interactions with the electroweak bosons, with a coupling strength suppressed by the mixing angles. 
Due to its gauge singlet nature, a Majorana mass term for $N$ may be present in the theory, in which case $N$ will be a Majorana particle. It is also possible to formulate models with Dirac or pseudo-Dirac HNLs. 
For the purposes of characterizing experimental sensitivities, we will follow the common convention of considering a single HNL that dominantly mixes with a specific neutrino flavor, i.e., dominant electron-, muon-, or tau- flavor mixing. The phenomenology is then characterized by the HNL mass, $m_N$, and mixing angle, denoted by $|U_e|^2$, $|U_\mu|^2$, $|U_\tau|^2$, respectively, for the three mixing scenarios. 

\begin{figure}[tbh]
\begin{center}
\includegraphics[width=0.9\textwidth]{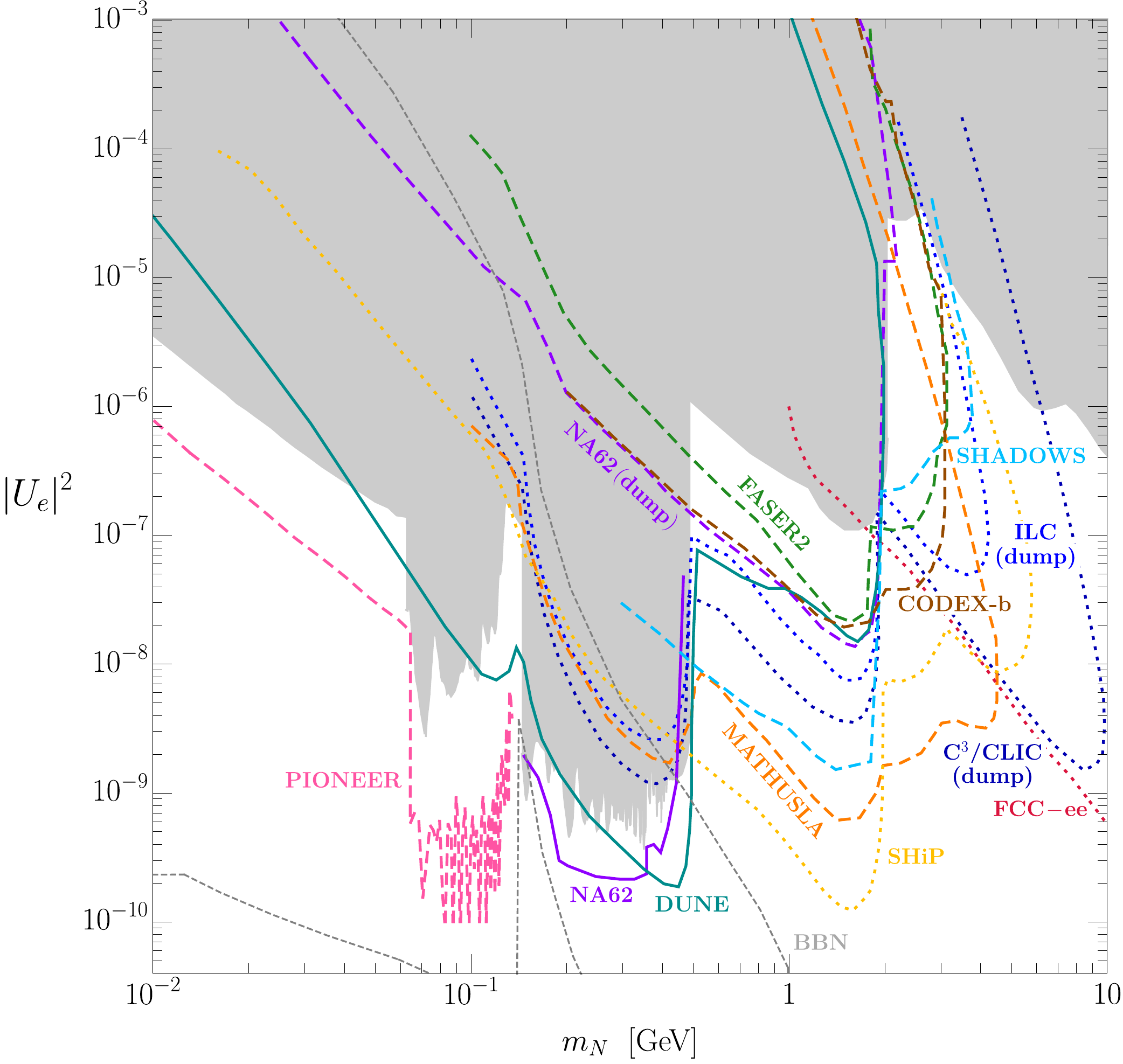}
\end{center}
\caption{
\textit{Neutrino portal with electron mixing dominance.}
Existing constraints (gray shaded region) and future experimental projections (colored contours) are shown in the $m_{N}-|U_e|^2$ plane. The existing bounds are from 
Refs.~\cite{Bryman:2019bjg,PIENU:2017wbj,NA62:2020mcv,T2K:2019jwa,Bernardi:1985ny,Bernardi:1987ek,CHARM:1985nku,Belle:2013ytx,DELPHI:1996qcc,CMS:2022fut,ATLAS:2022atq,ATLAS:2019kpx,Sabti:2020yrt}.
Also shown are projections from a number of existing and proposed future experiments, including 
DUNE~\cite{Berryman:2019dme,Ballett:2019bgd,Coloma:2020lgy},
NA62~\cite{NA62:2020mcv},
NA62-dump~\cite{Beacham:2019nyx},
FASER2~\cite{Kling:2018wct,Feng:2022inv},
FASER2~\cite{Kling:2018wct,Feng:2022inv},
CODEX-b~\cite{Aielli:2019ivi,Aielli:2022awh},
MATHUSLA~\cite{Curtin:2018mvb,MATHUSLA:2020uve},
SHADOWS~\cite{Baldini:2021hfw},
SHiP~\cite{SHiP:2018xqw}, 
ILC (250 GeV) and C3/CLIC (1000 GeV) beam dumps ~\cite{Giffin:2022rei}, and FCC-ee~\cite{Blondel:2014bra,Antusch:2016vyf,EuropeanStrategyforParticlePhysicsPreparatoryGroup:2019qin}.
}
\label{fig:HNLe}
\end{figure}

\begin{figure}[tbh]
\begin{center}
\includegraphics[width=0.9\textwidth]{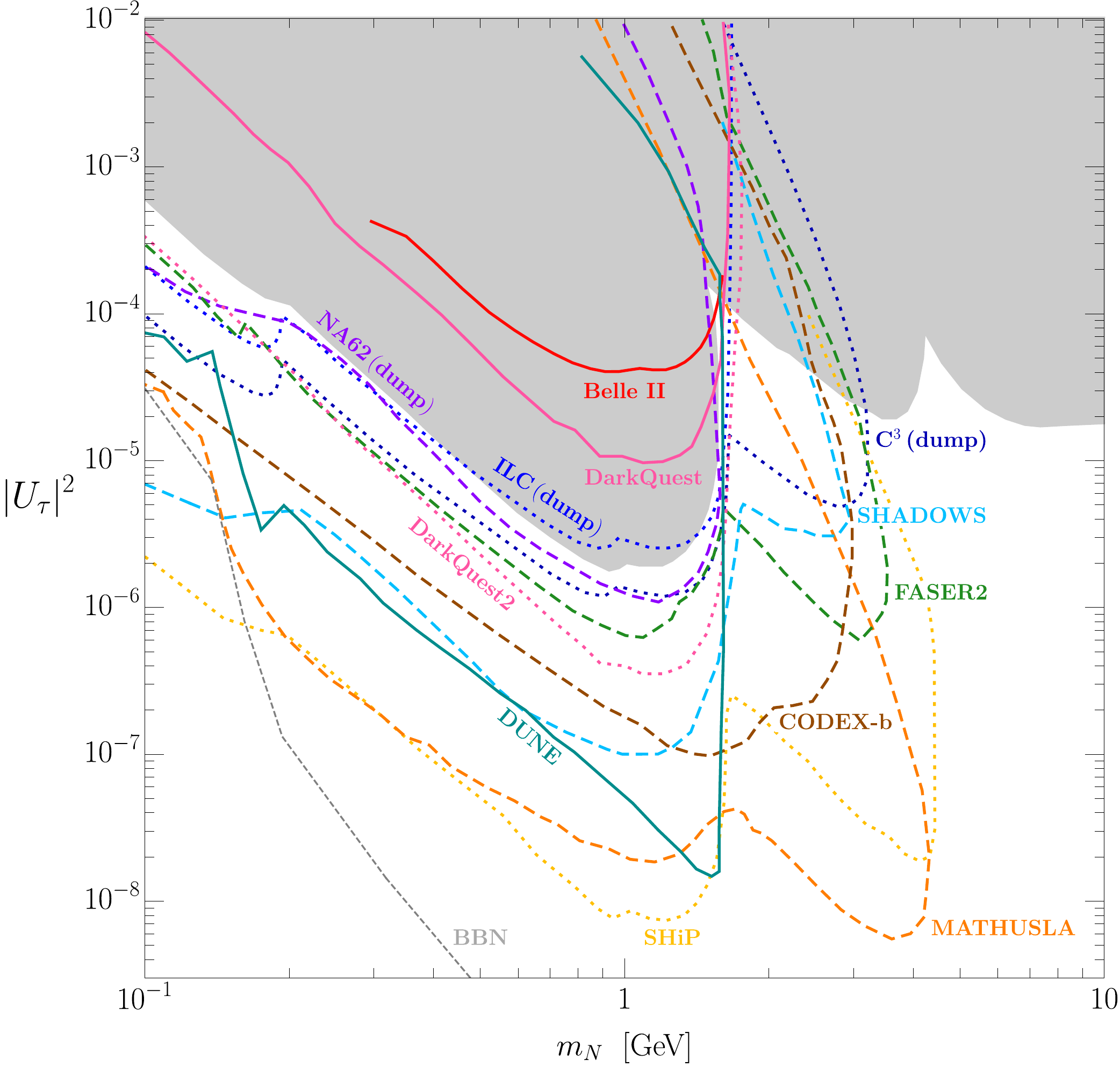}
\end{center}
\caption{
\textit{Neutrino portal with tau mixing dominance.}
Existing constraints (gray shaded region) and future experimental projections (colored contours) are shown in the $m_{N}-|U_\tau|^2$ plane.
The existing bounds are from 
Refs.~\cite{DELPHI:1996qcc,Boiarska:2021yho,CHARM:1985nku,Sabti:2020yrt}.
Also shown are projections from a number of existing and proposed future experiments, including 
DUNE~\cite{Berryman:2019dme,Ballett:2019bgd,Coloma:2020lgy},
DarkQuest~\cite{Batell:2020vqn,Berlin:2018pwi},
NA62~\cite{NA62:2020mcv},
NA62-dump~\cite{Beacham:2019nyx},
FASER2~\cite{Kling:2018wct,Feng:2022inv},
MATHUSLA~\cite{Curtin:2018mvb,MATHUSLA:2020uve},
CODEX-b~\cite{Aielli:2019ivi,Aielli:2022awh},
SHADOWS~\cite{Baldini:2021hfw}, SHiP~\cite{SHiP:2018xqw}, and ILC (250 GeV) and C3 (3000 GeV) beam dumps ~\cite{Giffin:2022rei}
}
\label{fig:HNLtau}
\end{figure}

Much like SM neutrinos, HNLs are readily produced through weak interaction decays, such as muon, pion, kaon, $D$ meson, tau lepton, and $B$ meson decays. HNLs also decay through the weak interactions and typically feature a wide variety of visible/semi-visible decay modes involving charged leptons, hadrons, and/or neutrinos. HNLs can thus be probed at a variety of experiments, including in dedicated pion or kaon decay experiments, beam dump experiments, and with a variety of experiments at the LHC. Figures~\ref{fig:HNLe} and \ref{fig:HNLtau} summarize the current bounds and future experimental sensitivities for the electron-dominance and tau-lepton-dominance scenarios, respectively.

HNLs are particularly well motivated given their likely connection with the generation of neutrino masses through the seesaw mechanisms~\cite{Minkowski:1977sc,Yanagida:1979as,GellMann:1980vs,Glashow:1979nm,Mohapatra:1979ia,Schechter:1980gr}. In addition to this broad motivation from the observation of neutrino masses, the neutrino portal could help explain the baryon asymmetry \cite{Asaka:2005pn} through the ARS leptogenesis mechanism \cite{Akhmedov:1998qx} and dark matter via a light keV-scale sterile neutrino~\cite{Boyarsky:2018tvu}. It has also been motivated by models of secluded DM \cite{Pospelov:2007mp}. It also exists in extensions of the SM which copy the SM itself, as in twin Higgs solutions to the little hierarchy problem. In order to achieve a consistent cosmology for some such models, the neutrino portal is necessary with masses and mixings of the sterile neutrinos detectable at near-future experiments \cite{Chacko:2016hvu}. 

Further discussion of the theoretical motivations and experimental prospects for HNLs can be found in the recent Snowmass whitepaper~\cite{Abdullahi:2022jlv}.

\subsection{ALP Portal}
\label{subsec:ALP-portal}

ALPs are parity-odd scalars that can arise as (pseudo-) Nambu-Goldstone bosons of spontaneously broken global symmetries or as compactifications of higher-dimensional gauge fields. These theoretical features ensure that ALPs can be very light compared to the scale of fundamental dynamics that generates them (e.g., the Peccei-Quinn breaking scale or the Planck scale). The QCD axion is the specific realization of the ALP that interacts with QCD (and gets its mass only from QCD effects which fix the mass as a function of the ALP coupling) and provides one of the most compelling explanations for the strong CP problem~\cite{Peccei:1977hh,Peccei:1977ur,Wilczek:1977pj,Weinberg:1977ma}. Here we consider the more general ALPs where the mass and coupling to SM particles are independent parameters. From an effective field theory point of view, ALPs couple to the SM through dimension five operators, suggesting that ALPs can be some of the first messengers of UV physics. As such, they also provide a compelling portal to dark sectors that require interactions with the SM~\cite{Batell:2009di}, such as certain models of DM (see, e.g., Ref.~\cite{Hochberg:2018rjs}).\footnote{For much lighter masses and weaker couplings than considered here, ALPs provide viable DM candidates themselves~\cite{Abbott:1982af,Dine:1982ah,Preskill:1982cy,Arias:2012az}.} It is therefore important to search broadly for the interactions of ALPs with SM particles. 

\begin{figure}[tbh]
\begin{center}
\includegraphics[width=1\textwidth]{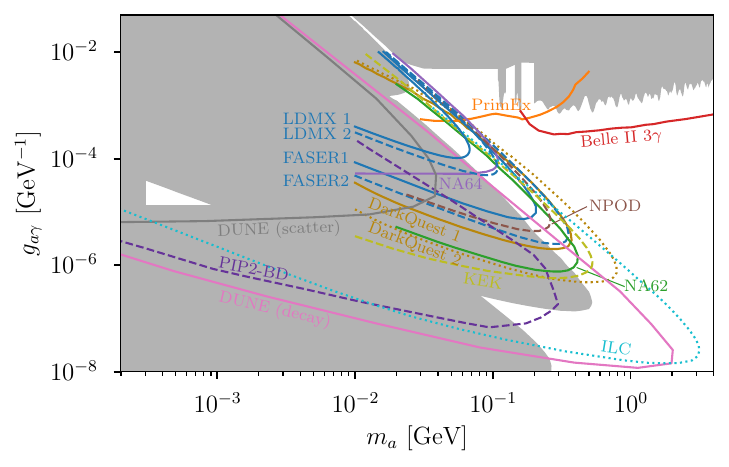}
\end{center}
\caption{\emph{Photon-coupled ALPs.} Existing constraints (gray) and projected sensitivities (colored lines) of various experiments to the photon-coupled ALP. The constraints are obtained from Refs.~\cite{Dolan:2017osp,Dobrich:2019dxc,Jaeckel:2015jla,BelleII:2020fag,Aloni:2019ruo,Lucente:2020whw,Ertas:2020xcc}. We show the sensitivities of phase 1 and 2 of DarkQuest~\cite{Berlin:2018pwi,Blinov:2021say, Apyan:2022tsd}; phase 1 and 2 of LDMX (visible search mode)~\cite{Berlin:2018bsc}; PIP2-BD~\cite{Toups:2022yxs}; phase 1 and 2 of FASER~\cite{Feng:2018pew,FASER:2018eoc};  NA62~\cite{Dobrich:2019dxc}; NA64~\cite{Dusaev:2020gxi}; phase 0 of LUXE-NPOD~\cite{Bai:2021dgm}; Belle II~\cite{Dolan:2017osp}; a reanalysis of existing PrimEx data~\cite{Aloni:2019ruo}; DUNE~\cite{Brdar:2020dpr}; KEK Linac~\cite{Ishikawa:2021qna}; and ILC beam dump~\cite{Sakaki:2020mqb}.}
\label{fig:alp_photon}
\end{figure}

\begin{figure}[tbh]
\begin{center}
\includegraphics[width=1\textwidth]{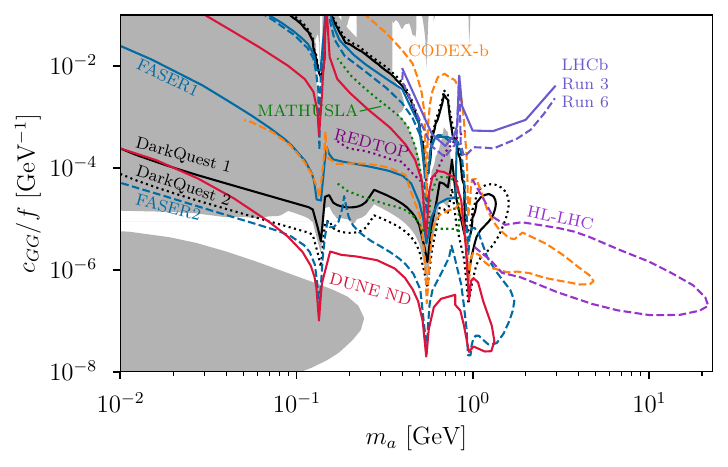}
\end{center}
\caption{\emph{Gluon-coupled ALPs (light-flavor only).} Existing constraints (gray) and projected sensitivities (colored lines) of various experiments to the gluon-coupled ALP, where ALP production is dominantly through light-flavor hadrons (protons, $\pi$, $K$, $\eta^{(')}$ mesons). The existing constraints are from Refs.~\cite{KOTO:2018dsc,NA62:2021zjw, NA62:2014ybm,ATLAS:2021kxv,Blumlein:1990ay,Blumlein:1991xh,CHARM:1985anb,Altmannshofer:2019yji,PIENU:2017wbj,Pocanic:2003pf,GlueX:2021myx,BelleII:2020fag,Jaeckel:2015jla,Ertas:2020xcc}. We show the projected sensitivities of phase 1 and 2 of DarkQuest~\cite{Blinov:2021say, Apyan:2022tsd}; phase 1 and 2 of FASER~\cite{FASER:2018eoc,Feng:2022inv}; CODEX-b~\cite{Aielli:2019ivi}; REDTOP~\cite{Beacham:2019nyx}; MATHUSLA~\cite{Chou:2016lxi,Beacham:2019nyx}; DUNE~\cite{Kelly:2020dda}; LHCb~\cite{Craik:2022riw} and HL-LHC Track Trigger~\cite{Hook:2019qoh,Kelly:2020dda}. 
We emphasize that these projections and sensitivities are for an ALP model where the gluon coupling is the only relevant interaction at low energies, which is a somewhat contrived assumption. If the interaction is instead defined at a high scale (like a TeV), renormalization group evolution generates a multitude of other interactions leading to additional constraints and discovery opportunities as illustrated in Fig.~\ref{fig:alp_gluon_uv}.
}
\label{fig:alp_gluon}
\end{figure}

\begin{figure}[tbh]
\begin{center}
\includegraphics[width=1\textwidth]{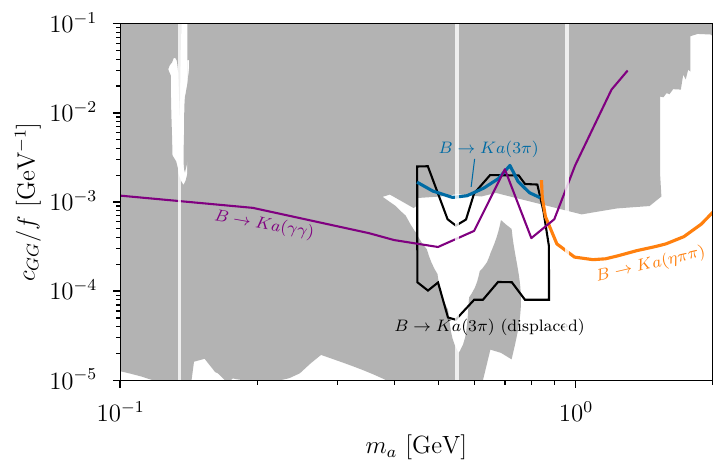}
\end{center}
\caption{\emph{Gluon-coupled ALPs.} Existing constraints (gray) and projected sensitivities (colored lines) of various experiments to the gluon-coupled ALP. Compared to Fig.~\ref{fig:alp_gluon}, in this figure we include interactions with heavy flavor quarks and with leptons, which are generated through renormalization group evolution from a high scale (their natural size can also be estimated from the finite pieces of counter-terms required to define the theory at low scales). The new constraints arise from searches for $B\to K a(\mu^+\mu^-)$, $\Upsilon \to \gamma a(\mathrm{hadrons})$~\cite{Bauer:2021mvw}, and 
$b\to s a$, $B\to K a (\gamma\gamma)$, $B\to Ka (3\pi)$, $B\to K a(KK \pi)$ and $B\to K a (\phi\phi)$~\cite{Chakraborty:2021wda,Bertholet:2021hjl}.
The projected sensitivities of searches for 
rare $B$ decays $B\to K a (3\pi)$ (prompt and displaced), $B\to K a(\eta\pi\pi)$ and $B\to K a (\gamma\gamma)$ at Belle II~\cite{Chakraborty:2021wda,Bertholet:2021hjl} are shown by solid lines. All of the projections from Fig.~\ref{fig:alp_gluon} also apply here, but we do not show them for clarity.
The vertical white lines at $m_a \approx m_{\pi^0}$, $m_{\eta}$ and $m_{\eta'}$ indicate parameter space where the ALP-meson mixing approximation used to interpret experimental results in terms of the gluon coupling is unreliable. 
}
\label{fig:alp_gluon_uv}
\end{figure}

ALP interactions depend strongly on the UV physics from which they arise. As a result, there is a wide variety of dimension five operators that can be manifest at experimental energies. Here we focus on two simple, minimal examples that capture a broad (but incomplete) range of possible production and decay modes. We will consider models where the ALP couples dominantly to photons or to gluons just above the QCD scale; the corresponding interactions are given by 
\beq
\mathcal{L}\supset c_{\gamma\gamma}\frac{\alpha}{4\pi} \frac{a}{f}F_{\mu\nu} \widetilde{F}^{\mu\nu} + c_{GG}\frac{\alpha_s}{4\pi} \frac{a}{f} G_{\mu\nu}^a \widetilde{G}^{a,\,\mu\nu},
\label{eq:alp_lagrangian}
\eeq
where $a$ is the ALP field, $F$ ($G$) are the electromagnetic (gluon) field-strength tensors, and we take only one of $c_{\gamma\gamma}/f$ or $c_{GG}/f$ to be non-zero at a time (these coupling constants have inverse mass dimension). In the gluon-coupled case, QCD confinement generates interactions of the ALP with hadrons, including mixing with pseudo-scalar mesons, which in turn couple to photons. Thus, in both the photon- \emph{and} gluon-coupled ALP scenarios, the ALP interacts with photons. This enables a very broad search strategy for ALPs of both kinds in precision-frontier experiments. ALPs are produced in collisions of SM particles via their couplings to photons or hadrons and travel a macroscopic distance before decaying into pairs of photons (in the gluon-coupled case, non-photon decay modes are also available). In Figs.~\ref{fig:alp_photon},~\ref{fig:alp_gluon} and~\ref{fig:alp_gluon_uv}, we show existing constraints and projections for future searches in the photon- or gluon-coupled ALP models. 

It is important to note that low-energy interactions of the gluon-coupled ALP are still somewhat uncertain because they involve strongly-coupled physics of QCD below the confinement scale. Several theoretical and data-driven models have been proposed~\cite{Aloni:2018vki,Bauer:2020jbp,Bauer:2021wjo}. In particular, Refs.~\cite{Bauer:2020jbp,Bauer:2021wjo} emphasized that many existing hadronic ALP calculations suffer from certain theoretical inconsistencies that can significantly impact the interpretation of experimental results. While it is unlikely that these issues lead to qualitative changes of the various experimental constraints and projections discussed below, it will be important to resolve this theoretical issue in the future. 

As mentioned above, the ALP interactions are sensitive to their UV origin; we specialized to two models that are minimal at low energies, featuring only a single defining coupling just above the QCD scale. This is a simple model in practice, but somewhat contrived from a theoretical point of view, as the non-renormalizable ALP interaction requires the introduction of counterterms to cancel off divergences in loop diagrams~\cite{Chakraborty:2021wda}. Thus the models with a single coupling just above the QCD scale require a careful fine-tuning of many operators to nullify their effects. The parameter space for this fine-tuned scenario is shown in Fig.~\ref{fig:alp_gluon}.

A different approach to studying the ALP model is to begin with a UV complete model at a high scale (say, a TeV), integrate out heavy physics and then evolve the resulting interactions to experimental scales~\cite{Bauer:2021mvw}. This procedure generically produces many interactions in addition to the gluon and photon couplings in Eq.~\ref{eq:alp_lagrangian}. These other interactions can provide complementary constraints~\cite{Bauer:2021mvw} and discovery opportunities~\cite{Chakraborty:2021wda,Bertholet:2021hjl}. The parameter space that is most affected by these additional constraints and projections is shown in Fig.~\ref{fig:alp_gluon_uv} (all other constraints and projections from Fig.~\ref{fig:alp_gluon} are still relevant here as well).

With these caveats in mind, in Fig.~\ref{fig:alp_photon} we show the existing constraints and projections for the photon-coupled ALP model. The constraints, shaded in gray, come from old beam dump experiments (E137 and E141~\cite{Dolan:2017osp}, CHARM and $\nu$CAL~\cite{Dobrich:2019dxc}); colliders (LEP~\cite{Jaeckel:2015jla}, BaBar~\cite{Dolan:2017osp} and Belle II~\cite{BelleII:2020fag}); the photon beam experiment PrimEx~\cite{Aloni:2019ruo}; supernova 1987A~\cite{Lucente:2020whw} and big bang nucleosynthesis~\cite{Ertas:2020xcc}.\footnote{We show the least stringent constraint derived by allowing $\Delta N_{\mathrm{eff}}$ and neutrino chemical potential to vary.} We also show the projected sensitivities for the following experiments: phase 1 and 2 of DarkQuest~\cite{Berlin:2018pwi,Blinov:2021say, Apyan:2022tsd}, PIP2-BD~\cite{Toups:2022yxs}, phase 1 and 2 of FASER~\cite{Feng:2018pew,FASER:2018eoc}; phase 1 and 2 of LDMX (search for visible energy depositions in the HCAL)~\cite{Berlin:2018bsc}, NA62~\cite{Dobrich:2019dxc}; NA64~\cite{Dusaev:2020gxi}; phase 0 of LUXE-NPOD~\cite{Bai:2021dgm}; Belle II~\cite{Dolan:2017osp}; a reanalysis of existing PrimEx data~\cite{Aloni:2019ruo}; DUNE~\cite{Brdar:2020dpr} and the ILC beam dump~\cite{Sakaki:2020mqb}.

In Fig.~\ref{fig:alp_gluon} we show the existing constraints and future projections for the gluon-coupled ALP scenario with only the ALP-gluon coupling just above the QCD scale (this is a simple but contrived scenario as mentioned above, in which the ALP is dominantly produced from light-flavor hadrons). The constraints, shaded in gray, come from KOTO~\cite{KOTO:2018dsc}; NA62~\cite{NA62:2021zjw, NA62:2014ybm}; ATLAS~\cite{ATLAS:2021kxv}; $\nu$CAL~\cite{Blumlein:1990ay,Blumlein:1991xh} and CHARM~\cite{CHARM:1985anb}; PIENU and PIBETA~\cite{Altmannshofer:2019yji, PIENU:2017wbj,Pocanic:2003pf}; GlueX~\cite{GlueX:2021myx}; recasts of the photon-only bounds from Belle II~\cite{BelleII:2020fag} and LEP~\cite{Jaeckel:2015jla}; and supernova 1987A~\cite{Ertas:2020xcc}. We also show the projected sensitivities of phase 1 and 2 of DarkQuest~\cite{Blinov:2021say, Apyan:2022tsd}; phase 1 and 2 of FASER~\cite{FASER:2018eoc,Feng:2022inv}; CODEX-b~\cite{Aielli:2019ivi}; REDTOP~\cite{Beacham:2019nyx}; MATHUSLA~\cite{Chou:2016lxi,Beacham:2019nyx}; DUNE~\cite{Kelly:2020dda}; LHCb~\cite{Craik:2022riw}; and HL-LHC Track Trigger~\cite{Hook:2019qoh,Kelly:2020dda}.

From a theoretical point of view it is much more reasonable to define the ALP-gluon coupling at a high scale corresponding to the scale at which beyond-SM matter is integrated out. As mentioned above, evolution from this high scale generates new interactions of the ALP that lead to additional constraints and discovery opportunities, beyond those shown in Fig.~\ref{fig:alp_gluon}. 
The natural size of the corresponding Wilson coefficients can be either computed explicitly using renormalization group evolution~\cite{Bauer:2020jbp,Chakraborty:2021wda}, or estimated by using these as counter-terms to cancel divergences in loop diagrams~\cite{Chakraborty:2021wda}.
Among these interactions, couplings to heavy flavor quarks and leptons are experimentally important. In Fig.~\ref{fig:alp_gluon_uv} we show the ALP parameter space which, in addition to all of the constraints from Fig.~\ref{fig:alp_gluon}, also includes bounds from LHCb ($B\to K a(\mu^+\mu^-)$), BaBar ($\Upsilon \to \gamma a(\mathrm{hadrons})$) as obtained in \cite{Bauer:2021mvw} (for a UV scale of $\sim 12$ TeV), and 
$b\to s a$, Belle ($B\to Ka (3\pi)$) and BaBar ($B\to K a (\gamma\gamma)$, $B\to K a(KK \pi)$, and $B\to K a (\phi\phi)$) as calculated in ~\cite{Chakraborty:2021wda,Bertholet:2021hjl} (for a UV scale of $\sim 1$ TeV), as well as projections from $B\to K a (3\pi)$ (prompt and displaced), $B\to K a(\eta\pi\pi)$ and $B\to K a (\gamma\gamma)$ at Belle II~\cite{Chakraborty:2021wda,Bertholet:2021hjl}.
Note that all of the projections from Fig.~\ref{fig:alp_gluon} also apply to Fig.~\ref{fig:alp_gluon_uv}, but we do not display them for clarity.

The coupling of ALPs to photons is predicted at near-detectable levels in many proposed answers to the big questions. A special, pion-phobic ALP which has predictive couplings to photons may be the long sought-after QCD axion \cite{Alves:2017avw}. Such a QCD axion could even help explain some nuclear de-excitation anomalies in \ce{^{8}Be} and \ce{^{4}He} \cite{Alves:2020xhf}. A heavier QCD axion may instead ammeliorate the ``axion quality problem'' of traditional QCD axion models, with a finite parameter space that will continue to be explored by future ALP searches \cite{Hook:2019qoh}.

ALPs also appear in solutions to other outstanding big questions besides the Strong CP problem, the original motivation for QCD axions. For example, coupling ALPs to both photons and dark-sector particles can provide a portal from DM to the SM. This ALP portal may provide the requisite thermalization between the SM and dark sectors during freeze-out in a SIMP DM model \cite{Hochberg:2018rjs}, with almost the entire viable parameter space observable in ALP searches at near-future experiments. If instead semi-annihilations of DM to ALPs set the relic abundance, the distinct parameter space which reproduces the measured DM abundance still predicts some benchmarks detectable at future ALP searches \cite{Kamada:2017tsq}.

\section{Outlook}
\label{sec.Outlook}

The minimal portals featured in this whitepaper are of intrinsic theoretical interest, constituting an essential ingredient in a variety of motivated dark sector theories that address some of the basic mysteries in particle physics and cosmology. 
Intensity frontier experiments have great potential to explore the renormalizable vector, Higgs, and neutrino portals, as well as the minimal ALP-gluon and ALP-photon portals, in the coming years. A variety of experimental facilities and detection strategies are required to probe the broadest range of mediator masses and portal couplings. These studies form a core part of a broader dark sector science program that also covers searches for dark matter production and rich dark sectors. 
Achieving the ambitious goals of this program necessitates investment in a variety of facilities and experiments, as well as support for theory research. 
Building on an already distinguished dark sector research tradition, there is great opportunity for U.S. scientists and institutions to provide key leadership in this enterprise over the next decade and beyond. 

\vspace{20pt}

{\bf Acknowledgements.}~B.~B.~is supported by U.S. Department of Energy (DOE) Grant DE–SC0007914 and by PITT PACC. 
N.~B.~was supported in part by NSERC, Canada.   
C.~H.~ is supported by the Natural Sciences and Engineering Research Council of Canada and Compute Canada.
R.~M.~is supported in part by the DoE grant DE-SC0007859.

\bibliography{bigidea2}

\end{document}